\documentclass[aps,prc,groupedaddress,twocolumn]{revtex4-1}
\usepackage{graphicx}
\usepackage[breaklinks]{hyperref}
\usepackage[english]{babel}
\usepackage{booktabs}
\usepackage{rotating}
\usepackage{multirow}
\usepackage{amsmath}

\begin{document}

\title{Comment on \textit{A dark matter interpretation of excesses in multiple direct detection experiments}}

\author{Alan E. Robinson and \'Emile Michaud}
\affiliation{D\'epartement de physique, Universit\'e de Montr\'eal, Montr\'eal, Canada H3C 3J7}
\email{alan.robinson@umontreal.ca}
\email{emile.michaud@umontreal.ca}

\date{February 20, 2020}

\begin{abstract}
In their recent preprint, Kurinsky, Baxter, Kahn, and Krnjaic assume an unphysical ionization yield for plasmon excitations in order to claim a possible dark matter signal.  Their proposed signal is not possible based on known physics, but their proposed detection method warrants further investigation.
\end{abstract}


\maketitle

In their recent preprint, Kurinsky, Baxter, Kahn, and Krnjaic \cite{plasmon} postulate that recent excesses in a variety of dark matter experiments searching for nuclear recoils and electron recoils may be produced by enhanced coupling of dark matter to plasmon resonances in these detector materials. Plasmons are a coherent excitation between electrons and ions that have been well studied in electron transmission \cite{raether} and inelastic x-ray scattering physics \cite{inelastic}.  They are expected to be seen in the spectra of eV-sensitive calorimeters exposed to keV and MeV-energy photons \cite{Emile}.

Section II C and Figure 2 of Kurinsky \textit{et al} compare excesses in two spectra of two runs of the EDELWEISS experiment.   The energy scale of these spectra are different linear combinations of total recoil energy and ionization energy sensitivity.  By choosing an ionization yield of 0.25 electron-hole pairs per 16 eV plasmon generated, they are able to support their dark matter interpretation.  As seen in electron energy loss spectroscopy \cite{EELS}, the dominant energy loss mechanism for ionizing electrons is plasmon excitations.  Thus, plasmon ionization is merely an intermediate step of electron ionization, and their ionization yields must nearly identical.  This is measured to be 1 electron-hole pair per 3.0 eV in germanium.  Additionally, regardless of the proportion of energy deposited in plasmons and other electronic excitations, similar ionization yields should result \cite{intermediate}.

As the authors mention, with a large electron yield per plasmon, it is difficult to interpret any of the observed excess signals as dark matter in light of the strong constraint from DAMIC for excesses of multi-electron/hole events \cite{Damic}.

The authors also propose in their Section IV A that secondary plasmons could be produced from high-momentum transfer collisions in association with phonons.  The dielectric function above the plasmon cutoff momentum is nearly constant in energy with no real resonances \cite{inelastic, Emile}.  Any excitation in this high-momentum high-energy regime should be expected to produce single electron states \cite{raether}.  Even if the phonon plus plasmon final state were probable, the phonon's momentum would lie well outside the first Brillouin zone.  The momentum required to extract 16 eV from  dark matter travelling near the galactic escape velocity is approximately 16~eV$ / 0.002c = 3.5 (2.27\text{ eV}/c)$ where 2.27~eV$/c$ is the inverse lattice spacing of silicon.  It should be possible to produce resonant plasmons directly in higher order Brillouin zones rather than relying on phonons to provide the required kinematic matching.

Kurinsky \textit{et al} conclude by noting the potential importance of collective effects in enhancing dark matter scattering with low-momentum transfer.  It should be noted that there are several models for dark matter with dilute charge distributions that would require such an significant form factor enhancement at low momentum transfers.  In particular, the axion-quark nugget model proposes composite dark matter particles with sizes of $\mathcal{O}(10^{-5})$~cm, resulting in a charge form factor that is suppressed at momenta $\gtrsim2$eV$/c$.  For such particles, Umklapp plasmon excitations may be a dominant energy loss mechanism.

The result of Kurinsky \textit{et al} \cite{plasmon} should not be taken as evidence for dark matter, although it does highlight the ongoing need to investigate the effect of collective modes how we detect radiation.

\begin{acknowledgments}
This research was undertaken thanks to funding from the Canada First Research Excellence Fund through the Arthur B. McDonald Institute.
\end{acknowledgments}

\end{document}